\documentclass[aps,prc,groupedaddress,twocolumn,showpacs]{revtex4-1}

\usepackage{amsmath,amssymb,bm}
\usepackage{ae}

\newcommand{\frc}[2]{\mbox{$\frac{#1}{#2}$}}

\newcommand{\pp}[1]{%
}

\usepackage{graphicx} 

\DeclareMathSymbol{\NS}{\mathord}{AMSb}{"4E}
\DeclareMathOperator{\abrapar}{\big<}
\DeclareMathOperator{\aketpar}{\big>}

\newcommand{\ket}[1]{\ensuremath{\,|{#1}\rangle}}
\newcommand{\braket}[2]{\ensuremath{\langle{#1}|{#2}\rangle}}
\newcommand{\matrixe}[3]{\ensuremath{\langle{#1}|\,{#2}\,|{#3}\rangle}}
\newcommand{\rmatrixe}[3]{\ensuremath{ \abrapar {#1} \big|\big| \,{#2}\, \big|\big| {#3} \aketpar }}

\newcommand{\expect}[1]{\ensuremath{\langle{#1}\rangle}}

\newcommand{\comm}[2]{\ensuremath{[{#1},{#2}]}}

\newcommand{\op}[1]{\ensuremath{#1}}
\newcommand{\adj}[1]{\ensuremath{{{#1}}^{\dag}}}

\renewcommand{\vec}[1]{\ensuremath{\bm{#1}}}

\newcommand{\cO}{\ensuremath{\op{c}}}

\newcommand{\rO}{\ensuremath{\op{r}}}

\newcommand{\alphaO}{\ensuremath{\op{\alpha}}}

\newcommand{\ccO}{\ensuremath{\adj{\op{c}}}}

\newcommand{\aalphaO}{\ensuremath{\adj{\op{\alpha}}}}

\newcommand{\AO}{\ensuremath{\op{A}}}

\newcommand{\HO}{\ensuremath{\op{H}}}

\newcommand{\QO}{\ensuremath{\op{Q}}}

\newcommand{\OOO}{\ensuremath{\adj{\op{O}}}}

\begin{document}


\title{Low-energy electric dipole response of Sn isotopes}

\author{P.~Papakonstantinou} 
\email[]{ppapakon@ibs.re.kr}
\affiliation{Institut de Physique Nucl\'eaire, IN2P3-CNRS, Universit\'e Paris-Sud, F-91406 Orsay CEDEX, France}
\affiliation{Rare Isotope Science Project, Institute for Basic Science, Daejeon 305-811, Korea}

\author{H.~Hergert}
\affiliation{Department of Physics, Ohio State University, Columbus, Ohio 43210, USA}

\author{V.Yu.~Ponomarev}
\affiliation{Institut f\"ur Kernphysik, Technische Universit\"at Darmstadt, 64283 Darmstadt, Germany}

\author{R.~Roth}
\affiliation{Institut f\"ur Kernphysik, Technische Universit\"at Darmstadt, 64283 Darmstadt, Germany}

\date{\today}

\begin{abstract}
%
%
%
We study the low-energy dipole (LED) strength distribution along the Sn isotopic chain in both the isoscalar (IS) and the isovector (IV, or E1) electric channels, to provide testable predictions and guidance for new experiments 
with stable targets and radioactive beams.
%
We use the self-consistent Quasi-particle Random-Phase Approximation (QRPA) with finite-range interactions and  mainly the Gogny D1S force. 
We analyze also the performance of a realistic two-body interaction supplemented by a phenomenological three-body contact term. 
%
We find that from $N=50$ and up to the $N=82$ shell closure ($^{132}$Sn) the lowest-energy part of the IS-LED spectrum is dominated by a collective transition whose properties vary smoothly with neutron number and which cannot be interpreted as a neutron-skin oscillation. 
For the neutron-rich species this state contributes to the E1 strength below particle threshold, 
but much more E1  
strength is carried by other, weak but numerous transitions around or above threshold. 
We find that strong structural changes in the spectrum take effect beyond $N=82$, namely increased LED strength and lower excitation energies. 
%
Our results with the Gogny interaction are compatible with existing data. 
On this basis we predict that (a) the summed IS strength  below particle threshold shall be of the same order of magnitude for $N=50-82$, 
(b) the summed E1 strength up to approximately 12~MeV shall be similar for $N=50-82$~MeV, while 
(c) 
the summed E1 strength below threshold shall be of the same order of magnitude for $N\approx 64 - 82$ and much weaker for the lighter, more-symmetric isotopes. 
We point out a general agreement of our results with other non-relativistic studies, 
the absence of a collective IS mode in some of those studies, 
and a possibly radical disagreement with relativistic models. 
\end{abstract}


\maketitle



\section{Introduction}

The literature on pygmy or soft dipole resonances~\cite{PVK2007,SAZ2013} of medium-mass and heavy nuclei has proliferated in the past two decades. 
The interest in such modes was triggered by possible relations of their properties with the symmetry energy of nuclear matter, neutron skins, and consequences of much astrophysical interest. 
Arguably, the meaning of such relations is compromised, when the very definition and existence of pygmy modes remains under debate. 
Nonetheless, it is widely accepted and can be safely considered a given, that non-negligible E1 strength is observed below particle threshold 
in many nuclei 
and that this strength cannot be explained (solely) as from the low-energy tail of the giant dipole resonance (GDR). 
In the following we will refer to this phenomenon merely as {\em low-energy dipole} (LED) strength, to avoid the term ``pygmy", which may mean different things to different readers.  
LED strength seems to be more significant in more-neutron rich nuclei. 
Merely the non-negligible amount of strength measured and the observation that it often appears in the form of at least one fragmented but resonant-like peak, especially in heavier nuclei, begs the question whether some kind of collective vibrational mode is at play. 
This is an interesting question for two reasons: 
(a) a collective mode would be of interest in itself and it is intuitive to expect that its properties are related to basic properties of matter or of finite quantum systems; 
(b) if no special collective transition is at play, then the observed strength must be largely from single-particle transitions, roughly unperturbed and losing only part of their strength to the giant dipole resonance; such a scenario has already been demonstrated a long time ago~\cite{OHD1998}. 
In such a case, it seems more constructive to focus the discussion on the shell structure of neutron-rich nuclei, no less an interesting issue. 

When considering nuclei as finite-size collections of two types of nucleons (protons and neutrons), quite a few normal dipole modes can be conceived. 
One of them is the oscillation of neutron matter against proton matter, associated with the GDR.  
Another is the oscillation of a layer of outer nucleons against an inert core. 
Among such core-layer modes, an isoscalar (IS) version involving protons and neutrons on a roughly equal footing has only recently attracted attention~\cite{Bas2008,Urb2012,PPR2011,PHP2012}. 
The core-layer mode most widely studied 
is the one whereby an outer layer of (mostly) neutrons 
oscillates against an (approximately) isospin-saturated core~\footnote{Alternatively: the excitation of excess neutrons against a core in configuration space, rather than real space} 
in a neutron-rich nucleus. 
The existence of such a mode was postulated a long time ago~\cite{MDB1971,SIS1990,VIN1992}. 
Early microscopic studies predicted the existence of precisely such modes already in stable nuclei~\cite{PVK2007}. 
The monotonic increase of LED strength with neutron excess predicted by such studies and an approximately linear correlation with quantities like the neutron skin and the neutron separation energy seemed to point to a simple resolution to the problem of what the origin and significance of LED (also called pygmy) strength could be, namely the neutron skin mode (which {\em also} came to be known as ``pygmy"). 

However, accumulating experimental data could not be reconciled with such studies in a satisfactory way.  
Although demonstrations have been attempted, establishing a clear and model-independent correlation of the above mode's centroid energy and strength with the asymmetry or neutron skin of the nucleus has not been possible so far. 
That all quantities vary monotonically within a model is not correlation enough.  
The attempts are hindered in part by the different energy intervals in which dipole strength is measured or calculated in different nuclei and studies. 
For such purposes, more promising appears to be the dipole polarizability~\cite{ReN2010,Tam2011,Pie2011}, largely because this quantity is less sensitive to the cutoff imposed on its calculation. 
%

The following open issues deserve particular attention: 
(a) The LED strength is often overestimated by microscopic calculations predicting neutron-skin modes, the explanation being that not all strength could be measured; a characteristic example is offered by relativistic models which claim to predict correctly the threshold dipole strength of unstable Sn isotopes, but overestimate the LED stength of stable ones (below threshold) by a factor of 2 or more~\cite{PVK2007}. 
(b) Theoretical studies predict a decrease of the centroid energy of LED strength with increasing neutron number, in disagreement with measured strength~\cite{Tof2011}. 
(c) From all the E1 transitions detected via photon scattering, only the lower-energy ones are visible in alpha scattering, i.e., by an IS probe~\cite{Sav2006,End2010,SAZ2013,Der2013}. 
It has been argued that the higher-energy transitions belong to the tail of the isovector GDR, while the lower-energy ones are due to a neutron-skin oscillation. 
However, to establish experimentally such a tail extending from the GDR region to below threshold in a roughly smooth way remains a difficult task. 
We note that studies of the IS coherence of the low-energy mode revealed the important role of both proton and neutron configurations~\cite{Lan2009,GGC2011,YKB2012}, leading to destructive coherence in the E1 channel. 

In this work we address the above open issues by studying the Tin isotopic chain, 
where quite some data already exist~\cite{Gov1998,Adr2005,Oze2007,Kli2007,Uts2009,End2010,Tof2011}  and new measurements on exotic species are planned or underway. 
Therefore, the scope of our methods can be assessed and validated using existing data and we are able to proceed to predictions which could be tested in the near future.  
If our predictions 
are not borne by experiment, 
our results will help initiate necessary improvements in broadly used theoretical methods; 
if they are borne by experiment, they will signify important progress in our understanding of LED strength. 
We note that a variety of theoretical studies exist already of the dipole response function of Sn isotopes~\cite{Gov1998,OHD1998,VPR2001,SBC2004,PNV2005,Pie2006,TeE2006,TsL2008,Lan2009,AGK2011,INY201Xa,YKB2012,Urb2012,DaG2012,BFC2012,Kva2011b,Kva2013,ENI2013a,ReN2013,CDA2013,LRT2013}. 
They include semiclassical approaches, the relativistic and non-relativistic random-phase approximation, and extended methods including phonon coupling, while many of them focus only on $^{132}$Sn. 
Generally speaking, non-relativistic models tend to interpret the low-energy E1 strength as coming from unperturbed single-particle states, to a large extent, while relativistic models tend to predict a collective neutron-skin mode below or just above particle threshold. 
We will be comparing our results with the various studies where appropriate.  

The present article is organized as follows: 
In Sec.~\ref{sec:moc} our theoretical method is presented. 
In Sec.~\ref{sec:res} we first validate our methods, in terms of global properties of the Sn isotopes, quantities related to the GDR and measured LED strength, and then we further analyse our results and we make predictions. 
We conclude in Sec.~\ref{sec:con}.


\section{Method of calculation\label{sec:moc}} 

We employ the self-consistent Quasi-particle Random-Phase Approximation (QRPA), which, in the absence of pairing, reduces to the usual Random-Phase Approximation for closed-shell nuclei. Spherical symmetry is assumed. 
Our implementation has been discussed in detail in Ref.~\cite{HPR2011} and is presented here briefly for completeness. 
Our starting point is the Hamiltonian
\begin{equation}
  H  =  T_{\mathrm{int}}  + V_{NN} +V_{\rho}  ,
\end{equation}
where the intrinsic kinetic energy is defined as 
\begin{equation}\label{eq:def_Tint}
   T_{\mathrm{int}} = T-T_\text{cm}=\left(1-\frac{1}{A}\right)\sum_i\frac{\vec{p}^2_i}{2m} - \frac{1}{mA}\sum_{i<j}\vec{p}_i\cdot\vec{p}_j\, 
,
\end{equation}
in an obvious notation. 
The two-body potential $V_{NN}$ is of finite range and includes the Coulomb interaction, 
while 
\begin{equation} 
V_{\rho} = t_3(1+x_3)\delta(\vec{r})\rho^{\alpha}(\vec{R})
\end{equation} 
is a density-dependent contact interaction, 
with $\vec{r}$ the relative and $\vec{R}$ the center-of-mass position vector of the interacting nucleon pair. 

We formulate the QRPA in the canonical basis of the Hartree-Fock Bogoliubov (HFB) ground state. 
The HFB implementation is presented in Ref.~\cite{HeR2009}. 
Single-particle states are expanded in a harmonic-oscillator basis of 15 major shells. 
The length parameter $a_{\mathrm{HO}}$ of the harmonic oscillator is determined for each nucleus such that 
the ground state energy is minimized. 
 
Assuming spherical symmetry, the canonical basis states come in pairs $\{\ket{\mu, m_\mu}, \ket{\overline{\mu, m_\mu}}\}$ which are related by time reversal:
\begin{equation}
  \ket{\overline{\mu m}} = (-1)^{l+j-m}\ket{\mu -m}\, .
\end{equation}
Here $\mu=(nlj\tau)$ indicates collectively the radial, angular momentum, and isospin quantum numbers. 
In the canonical basis, the Bogoliubov transformation between the particle and quasiparticle representation reduces to the BCS-like form 
\begin{subequations}\label{eq:def_qp}
  \begin{align}
    \aalphaO_{\mu m} &= u_{\mu} \ccO_{\mu m} + v_\mu \widetilde{\cO}_{\mu m}\,,\\
    \widetilde{\alphaO}_{\mu m} &= u_{\mu} \widetilde{\cO}_{\mu m} - v_\mu \ccO_{\mu m}\,,
  \end{align}
\end{subequations}
where the annihilation operators have been expressed as spherical tensors 
\begin{equation}\label{eq:def_sphadj}
   \widetilde{\alphaO}_{\mu m} = (-1)^{j+m}\alphaO_{\mu - m} = -(-1)^{l} \alphaO_{\overline{\mu m}}\,,
\end{equation}
and a factor $(-1)^{l}$ has been absorbed into the coefficients $v_\mu$ for simplicity.  

The QRPA phonon creation operator in the canonical basis has the general form 
\begin{equation}\label{eq:def_phonon_general}
   \OOO_k = \!\!\sum_{(\mu m) < (\mu' m')}\!\!X^k_{\mu m,\mu'm'}\aalphaO_{\mu m}\aalphaO_{\mu'm'} - Y^k_{\mu m,\mu'm'}\alphaO_{\mu'm'}\alphaO_{\mu m}
\,.
\end{equation} 
Spherical symmetry allows us to use an angular-momentum coupled representation 
\begin{equation}\label{eq:def_phonon_spherical}
   \OOO_{kJM} = \sum_{\mu \leq \mu'}X^{kJ}_{\mu\mu'}\mathcal{A}^\dag_{\mu\mu'JM} - Y^{kJ}_{\mu\mu'}\widetilde{\mathcal{A}}_{\mu\mu'JM}\,,
\end{equation}
where the coupled quasiparticle-pair creation operator is defined as
\begin{equation}
  \mathcal{A}^{\dag}_{\mu\mu'JM}\equiv\frac{1}{\sqrt{1+\delta_{\mu\mu'}}}\sum_{m,m'}
                     \braket{j m j'm'}{JM}\aalphaO_{\mu m}\aalphaO_{\mu'm'}
\end{equation}
and $\widetilde{\AO}_{\mu\mu'JM}$ is its spherical adjoint (cf. Eq. \eqref{eq:def_sphadj}).
Applying the Equations-of-Motion method 
we are able to define the QRPA matrices $A$ and $B$ via the commutators ($\mu\leq\mu',\nu\leq\nu'$)
\begin{subequations}\label{eq:def_AB}
  \begin{align}
    A^{JM}_{\mu\mu',\nu\nu'}&\equiv\matrixe{\Psi}{\comm{\widetilde{\mathcal{A}}_{\mu\mu'JM}}{\comm{\HO}{\mathcal{A}^\dag_{\nu\nu'JM}}}}{\Psi}\,,
    \\
    B^{JM}_{\mu\mu',\nu\nu'}&\equiv\matrixe{\Psi}{\comm{\widetilde{\mathcal{A}}_{\mu\mu'JM}}{\comm{\HO}{\widetilde{\mathcal{A}}_{\nu\nu'JM}}}}{\Psi}\,.
  \end{align}
\end{subequations}
Within the usual quasi-boson approximation, the ground state $\ket{\Psi}$ is the HFB vacuum. For spherically symmetric systems, the dependence on the angular momentum projection can be dropped and we obtain the reduced set of QRPA equations: 
\begin{equation}\label{eq:qrpa}
   \begin{pmatrix}
    A^J & B^J\\
   -B^{J*} & -A^{J*}
  \end{pmatrix}
  \begin{pmatrix}
   X^{kJ} \\ Y^{kJ}
  \end{pmatrix}
  = \hbar\omega_k
  \begin{pmatrix}
   X^{kJ} \\ Y^{kJ}
  \end{pmatrix}\,,
\end{equation}
where $\hbar\omega_k$ is the excitation energy of the $k$th QRPA state with respect to the ground state. 
The expressions for the matrices $A$ and $B$ are given in the Appendix of Ref.~\cite{HPR2011}. 
When constructing the QRPA equations, all possible two-quasiparticle configurations for the given $J^{\pi}$ 
are taken into account. 

For electric multipole transitions, the reduced transition probabilities are defined as
\begin{equation}\label{eq:def_bej}
   B(EJ, J_i\rightarrow J_f)\equiv\frac{1}{2J_i+1}\left|\rmatrixe{fJ_f}{\QO_J}{iJ_i}\right|^2\,.
\end{equation}
In the QRPA, we consider transitions from the $0^+$ ground state of an even-even nucleus to an excited state described by the QRPA phonon operator \eqref{eq:def_phonon_spherical}, and the reduced matrix element can be evaluated to 
\begin{align}\label{eq:def_transme}
  &\rmatrixe{kJ}{\QO_J}{0}\notag\\
&=\sum_{\mu\leq\mu'}\frac{1}{\sqrt{1+\delta_{\mu\mu'}}}\left(u_\mu v_{\mu'}+(-1)^Jv_\mu u_{\mu'}\right)\notag\\
  &\quad\times
       \left(X_{\mu\mu'}^{kJ*}\rmatrixe{\mu}{\QO_J}{\mu'}+
                  (-1)^JY^{kJ*}_{\mu\mu'}\rmatrixe{\mu}{\QO_J}{\mu'}^*\right)\,.
\end{align}
The IS and electric dipole response is determined, in the long-wavelength limit, by the following respective operators, 
which are corrected for spurious center-of-mass effects: 
\begin{equation}\label{eq:E1_IS_mod}
  \QO_{\mathrm{IS}}=e\sum_{i=1}^A\left(\rO_i^3-\frac{5}{3}\expect{\rO^2_\text{rms}}r_i\right)\sqrt{3}Y_{1M}(\hat{\vec{r}}_i)\,,
\end{equation}
and
\begin{equation}\label{eq:E1_IV_mod}
  \QO_{E1}=e\frac{N}{A}\sum_{p=1}^Zr_pY_{1M}(\hat{\vec{r}}_p)-e\frac{Z}{A}\sum_{n=1}^Nr_nY_{1M}(\hat{\vec{r}}_n)\,,
\end{equation}
where $\rO_\text{rms}$ is the root-mean-square radius operator. 
As far as intrinsic excitations are concerned, the operator $\QO_{E1}$ is equivalent to the uncorrected $E1$ operator and to the isovector dipole operator 
\[ 
 \QO_{E1} \equiv e\sum_{p=1}^Z r_pY_{1M}(\hat{\vec{r}}_p) \equiv \frc{e}{2}\sum_{i=1}^A \tau_3^{(i)} r_iY_{1M}(\hat{\vec{r}}_i)  
,
\]  
in an obvious notation. 
When we speak of {\em E1 strength} we will refer to the electric-dipole strength $B(E1)\uparrow$ corresponding to the operator $\QO_{E1}$ 
and we shall use {\em IS strength} when we refer to the operator $\QO_{\mathrm{IS}}$. 

Our QRPA implementation, which uses the same interaction in the ground states and $ph$ and $pp$ channels 
and avoids arbitrary truncations, 
achieves an excellent degree of self consistency, as discussed in Ref.~\cite{HPR2011}. 
The QRPA transition strengths obtained with the corrected operators (\ref{eq:E1_IS_mod}) and (\ref{eq:E1_IV_mod}) 
are the same as with the uncorrected forms, except of course for the spurious state. 
The spurious state appears always very close to zero energy, namely at most at a few tens of keV in the cases studied here. 

In order to assess the collectivity of a given state, single-particle units are useful indicators. 
In the case of E1 transitions the usual single-particle unit must be corrected for the center-of-mass motion, by multiplying the usual estimate~\cite{RS80} by the effective-charge factor squared, namely $(N/A)^2$ for proton states and $(Z/A)^2$ for neutron states~\cite{Hav2001}. 
As a reference quantity we will make use of the single-neutron unit 
\begin{equation} 
B_{\textrm{s.n.}}(E1)=0.06445A^{2/3}(Z/A)^2e^2\mathrm{fm}^2 
\label{eq:Bsn} 
\end{equation}  
 

In the present study we use the Gogny D1S interaction~\cite{BGG1991}, as well as a unitarily transformed Argonne V18 interaction supplemented with a phenomenological density-dependent two-body term~\cite{GRH2010,PPR2011,GPR20XX}. 
The transformation is achieved via the Unitary Correlation Operator Method (UCOM), 
with correlation operators determined 
through an evolution within the similarity normalisation group (SRG). 
The contact density-dependent term is derived from a corresponding contact three-body term (3N). We therefore have $\alpha =1$ in this case. The strength $t_3$ of the density-dependent term is determined such that ground-state energies are reproduced within perturbation theory, while $x_3$ is equal to 1. 
As in Ref.~\cite{PHP2012}, we will denote the transformed Argonne V18 interaction as 
UCOM(SRG)$_{S,\delta 3N}$.


\section{Results\label{sec:res}} 

\subsection{First comparisons with existing data\label{Subsec:GDR}} 

First we examine how well our theoretical method reproduces known properties of Sn isotopes. 
We may anticipate good results from Gogny D1S, which was developed to perform well in HFB calculations and has yielded 
interesting comparisons with data in Refs.~\cite{PPR2011,PHP2012}. 
We recall that 
UCOM(SRG)$_{S,\delta 3N}$, 
is meant to reproduce ground-state properties only within perturbation theory. 
Previous works~\cite{GRH2010,GPR20XX} have shown that within HFB it will underestimate the binding energies, 
while charge radii should be approximately correct, with a weak dependence on the choice of $a_{\mathrm{HO}}$. 
We also anticipate defects in the shell structure mainly because of the weak spin-orbit splittings~\cite{GRH2010}. 
The GDR is expected to be reproduced reasonably well with this interaction~\cite{Gue2011,GPR20XX}. 

\begin{figure}
\includegraphics[angle=-90,width=0.48\textwidth]{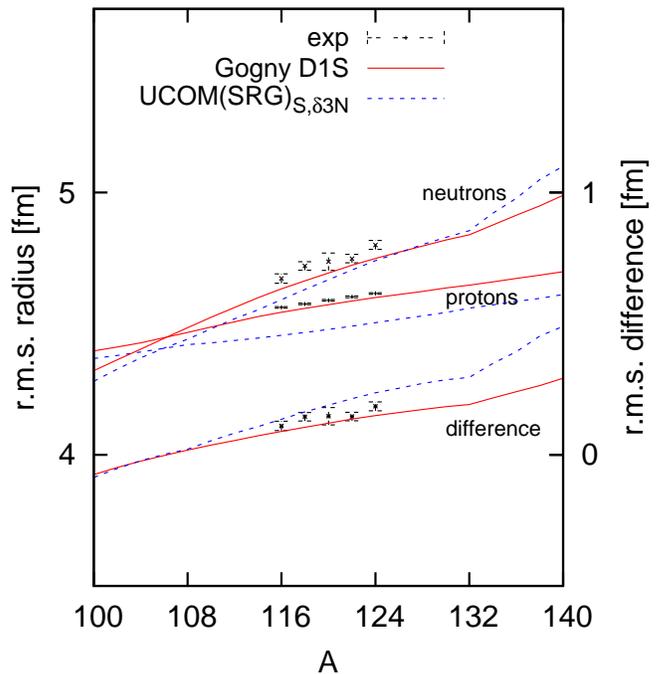}
\caption{(Color online) Calculated point-proton and neutron root-mean-square (r.m.s.) radii, 
as well as their differences, 
for the two indicated interactions, 
compared with measured r.m.s. radii~\cite{Ter2008}. 
\label{Fig:radii}}
\end{figure}

With the above in mind, we proceed to a comparison of HFB results for the two interactions with available data in Fig.~\ref{Fig:radii}. 
Neutron r.m.s. radii from elastic proton scattering on the even-even isotopes $^{116-124}$Sn were reported in Ref.~\cite{Ter2008}.  
Both interactions yield similar results for the neutron r.m.s., while  
UCOM(SRG)$_{S,\delta 3N}$ 
slightly underestimates the proton r.m.s. and consequently the charge radius. 
As a result, the corresponding values for the neutron skin, namely the difference between the proton and neutron r.m.s., 
are large compared to the Gogny D1S interaction, but roughly as compatible with the data. 
 
We proceed to the GDR region of the even-even isotopes $^{112-124}$Sn. 
In order to have a uniform comparison for all isotopes, we adopt the sets of 
photoneutron cross sections, $(\gamma ,xn)$, by Yu.I.Shorokin et al., compiled in~\cite{IAEA2000}. 
Cross sections were measured at excitation energies up to 27~MeV in all cases. 
The full width at half maximum, depending on the isotope, was found between 5.5 and 9~MeV in these experiments. 

\begin{figure}
\includegraphics[angle=-90,width=0.48\textwidth]{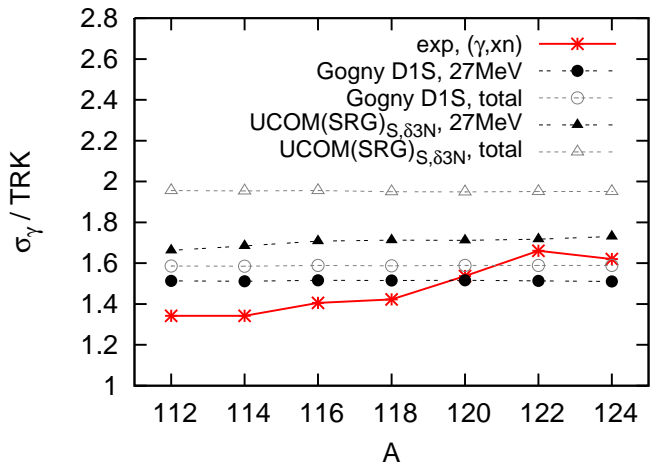}
\caption{(Color online) Calculated fraction of the classical Thomas-Reiche-Kuhn (TRK) sum rule, 
for the indicated isotopes and two interactions, 
up to 27~MeV or the whole excitation-energy region, 
compared with data obtained up to 27~MeV~\cite{IAEA2000}. 
 \label{Fig:totalE1}}
\end{figure}

In Fig.~\ref{Fig:totalE1} 
we show the portion of the classical TRK sum rule exhausted by the experimental data and by our theoretical calculations 
with both interactions. 
We have computed the TRK portion exhausted over all energies as well as only up to 27~MeV, to compare directly with data. 
The theoretical trends are more uniform than the experimental ones, 
but the overal quantitative comparison of the summed strength up to 27~MeV is very satisfactory. 
It is interesting to note that the 
UCOM(SRG)$_{S,\delta 3N}$ 
interaction predicts much more strength above 27~MeV than the Gogny interaction. 
However, the TRK fractions up to 27~MeV for the two interactions differ by only up to 15\%. 
 
\begin{figure*}
\includegraphics[angle=-90,width=\textwidth]{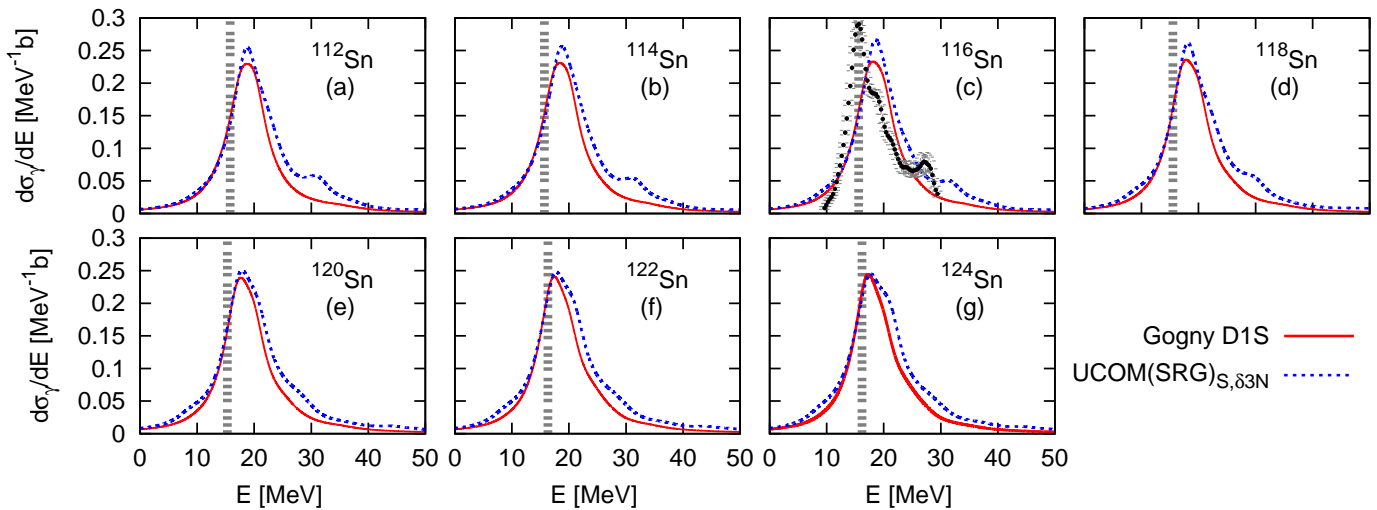}
\caption{(Color online) Calculated photoabsorption cross section, for the indicated interactions and all even isotopes 
for which data exist~\cite{CDFE}. 
The cross sections are smoothed with a Lorenzian of width equal to 5~MeV. 
The shaded vertical line indicates the experimental peak of the cross section. 
In the case of $^{116}$Sn the data are also shown.  \label{Fig:sigma}}
\end{figure*}

In Fig.~\ref{Fig:sigma} 
we plot the calculated photonuclear cross sections, with both interactions, smoothed with a lorenzian of width 5~MeV. 
The gray line indicates the peak energy of the measured cross section. 
Both interactions yield similar results up to the cross-section peak. 
They both overestimate its position. 
The UCOM(SRG)$_{S,\delta 3N}$ 
predicts systematically more strength at higher energies and in certain cases a secondary peak becomes obvious. 
Secondary peaks are present also in the experimental data. 
As an example, the measured cross section $(\gamma ,xn)$ is also shown in the case of $^{116}$Sn. 

Based on the above results, we conclude that the performance of both interactions is similar, as far as the GDR region is concerned. 
We also note the interesting result that the large enhancement factor corresponding to our quasi-realistic interaction 
does not entail disagreement with measured cross sections in the energy region of the GDR, but rather enhancement of strength at higher energies compared to other models. 

The low-energy strength and how it compares with data will be discussed throughout Sec.~\ref{Sec:LER}. 
The prominent role of shell structure at low energies will be apparent.  
 
\subsection{Properties of low-energy states\label{Sec:LER}} 

\subsubsection{General observations} 

\begin{figure}
\includegraphics[width=0.42\textwidth]{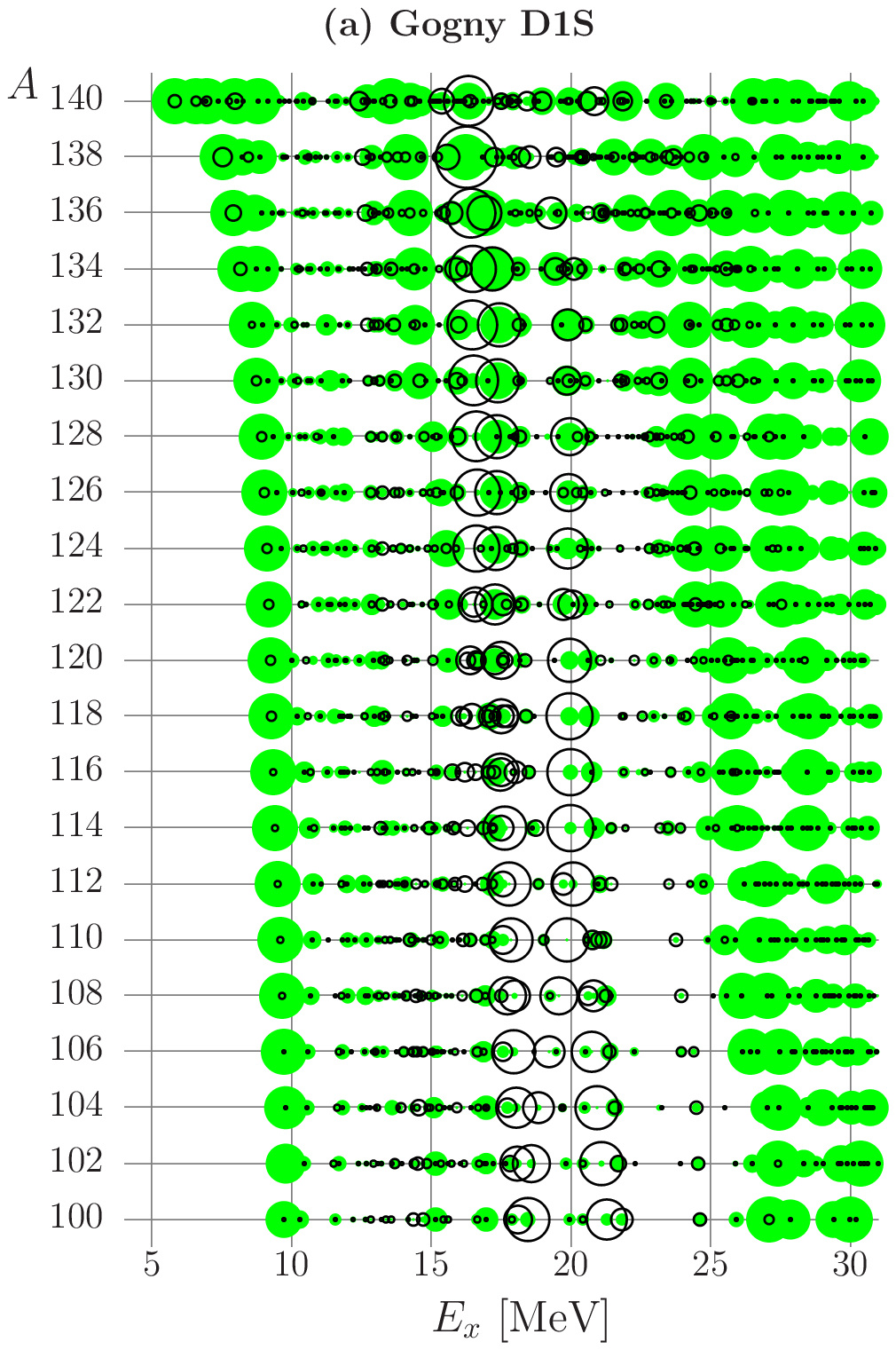}
\\[2mm] 
\includegraphics[width=0.42\textwidth]{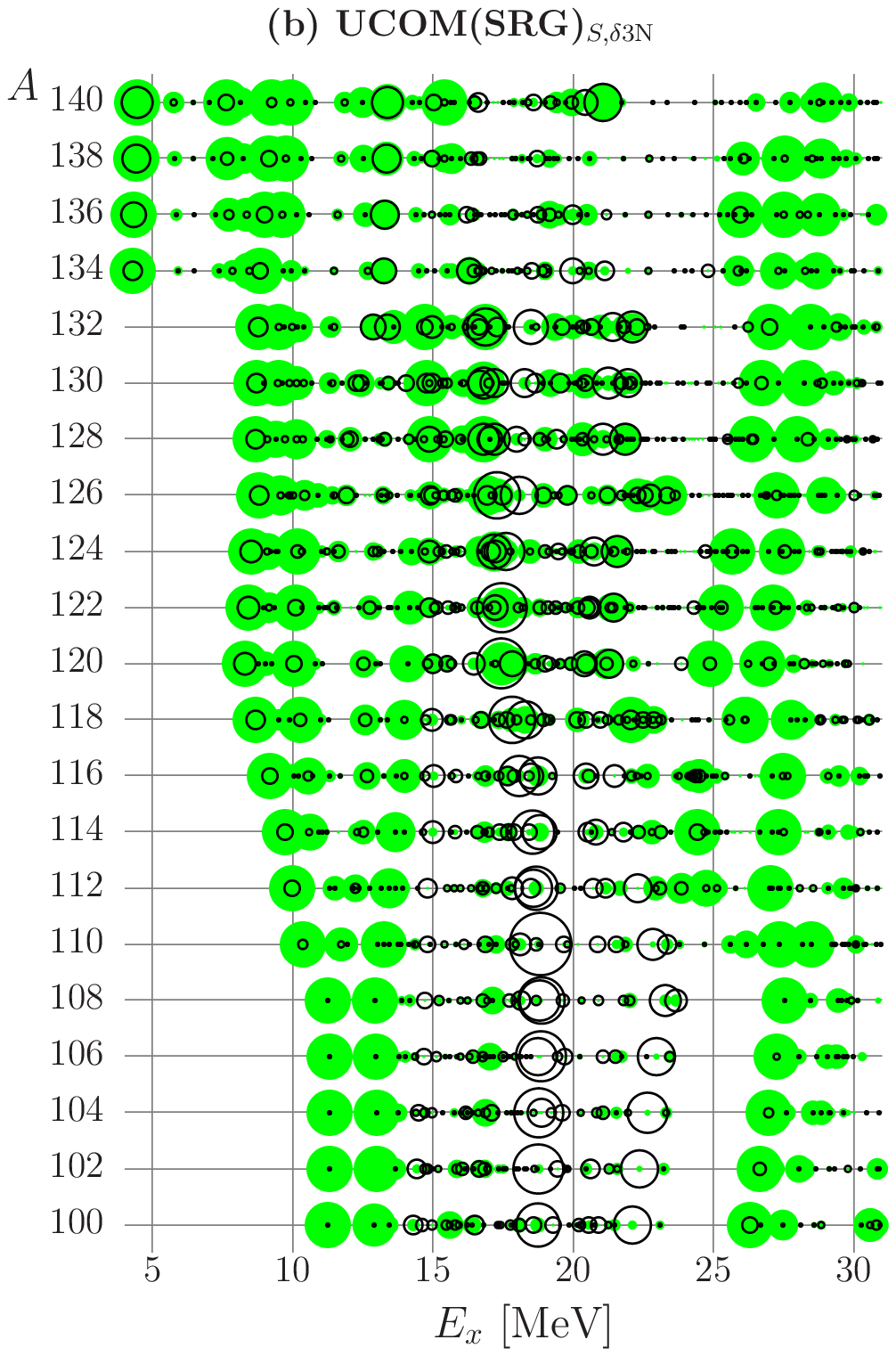}
\caption{
(Color online) For the given isotopes and excitation energies, 
the IS and E1 transition strengths  
are indicated by green disks and black circles, respectively. 
The area of a disk or circle is propotional to the strength. 
Results are given with the Gogny D1S and the UCOM(SRG)$_{S,\delta 3N}$ interactions. 
\label{Fig:Response}}
\end{figure}

Figure~\ref{Fig:Response} 
gives an overview of the response of all even Sn isotopes from $A=100$ to $A=140$, 
as calculated within QRPA with the two different interactions. 
The IS and E1 transition strengths  
are indicated by green disks and open black circles, respectively. 
The area of a disk or circle is propotional to the strength. 
The basic features of the spectrum and how they evolve with neutron number are clearly visible in this type of plot. 
The large circles in the energy region of $20\pm 3$~MeV correspond to the GDR. 
The GDR appears fragmented into at least two structures. 
The disks at high energy, above 25~MeV, correspond to the high-lying compression mode. 
Apart from these structures, we notice that the low-energy spectrum 
in the IS channel is dominated, in all isotopes, by a strong IS-LED transition at about 10~MeV. 
This result is in obvious qualitative agreement with the Skyrme-QRPA results of Ref.~\cite{TeE2006}. 
In the neutron-rich isotopes this transition appears to carry also considerable E1 strength. 
Important differences between the two interactions can be pointed out: 
For the Gogny D1S interaction, the energy and IS strength of the strong IS-LED state drops rather smoothly as the neutron number increases. 
Beyond $N=132$ the changes appear to accelerate. 
For the 
UCOM(SRG)$_{S,\delta 3N}$ 
the energy reaches a minimum at $A=120$ and then rises again up to $A=132$. 
The IS strength is more fragmented than in the case of Gogny D1S. 
From $A=110$ and upwards the IS-LED states carry apparently more E1 strength than in the case of Gogny D1S. 

In all cases, considerable (by comparison) and fragmented E1 strength appears between the major IS-LED state and the GDR. 
As in our similar study of Ca isotopes~\cite{PHP2012}, we note the rich structure of the response in the whole energy region up to the GDR. 

Many of the differences we observe between Fig.~\ref{Fig:Response}(a) and Fig.~\ref{Fig:Response}(b) at low energies are manifestations of the different shell structure predicted by the two interactions. The neutron single-particle energies, defined as the eigenvalues of the single-particle Hamiltonian, are quite relevant. 
Using the Gogny D1S interaction, we obtain a well defined energy gap of about 6~MeV between the intruder, valence state $\nu 0h_{11/2}$ and the lowest neutron $pfh$ state, namely $\nu 1f_{7/2}$. 
Thus $^{132}$Sn is a closed-shell nucleus, while $^{134}$Sn has $\nu 1f{7/2}$, its last bound neutron state, doubly occupied. 
By contrast, with the UCOM(SRG)$_{S,\delta 3N}$ interaction we obtain much weaker spin-orbit splittings, and as a result, $\nu 0h_{11/2}$ lies almost midway between the $sdf$ and the $pfh$ shells, and in fact closer to the latter. 
Due to this behavior, $^{132}$Sn is not a proper closed-shell nucleus, while pairing collapses mid-shell for $^{108}$Sn and $^{120}$Sn, namely the isotopes for which kinks are observed in the evolution of the lowest eigenenergies in Fig.~\ref{Fig:Response}(b). 
Another result obtained with the UCOM(SRG)$_{S,\delta 3N}$ interaction, is that the entire $pfh$ shell is unbound or very weakly bound (the same holds for $\nu 0h_{11/2}$ up to $A=108$). 
Thus the dramatic shift of strengths to lower energies for $A\geq 134$ signifies the activation of barely bound hole states.  
Shell defects such as the above have been discussed before~\cite{GRH2010}, but, as seen also in Sec.~\ref{Subsec:GDR}, they do not spoil the description of very collective phenomena like electric giant resonances~\cite{GPR20XX}. 

We will proceed to a few more comparisons between the two interactions and with data in Sec.~\ref{Subsec:FDS}, until quantitative arguments speak against the use of the above interaction in (Q)RPA studies of low-energy dipole response. 


In the following we will focus on excitation energies up to about 15~MeV. 
We expect the excitation energies to be overestimated by QRPA in this region, due to the lack of explicit long-range correlations and coupling to complex configurations and surface vibrations, as inferred from a variety of studies on different nuclei (see, e.g., \cite{CoB2001a,Har2004,GGC2011}).  
The overestimation of excitation energies can be avoided already at the mean field level by considering the nucleon effective mass $m^{\ast}$ approximately equal to the bare nucleon mass $m$. However, the in-medium effective mass resulting self-consistently from the present interactions is very low ($0.7m$ for Gogny D1S). 
As a result, the density of single-particle states close to the Fermi level is too low, 
the lowest $2qp$ energies in HFB are too high, on average, 
in comparison to phenomenological Woods-Saxon energies used, e.g., in \cite{Gov1998}, and the associated dipole strength appears too high by about 3 MeV.  
As a rule of thumb, we will therefore consider all energies overestimated by about 3~MeV in the following, as we did in Ref.~\cite{PHP2012}. 
Such a value is consistent also with the energetic discrepancies between (Q)RPA results and data observed in other studies (e.g., Ref.~\cite{GoK2002}).  
Of course, the closer we move to the GDR region, the more obsolete such a shift should be considered. 
To a good approximation, though, our discussion of results up to 15~MeV excitation energies should be relevant for an actual energy region up to, roughly, 12~MeV. 
This should be kept in mind in what follows. 
Another relevant observation is that the particle continuum and resonant particle states are not treated exactly in configuration-space RPA. 
It may be owing to this deficiency that, for example,  
the particle threshold of the stable isotopes, defined as the binding energy of the $0h_{11/2}$ neutron state, is found at around $7-8$~MeV with the Gogny force, but the first unperturbed $1^-$ $2qp$ transition to an unbound state, i.e., effectively the HFB particle threshold in this channel, appears well above 10~MeV. 

\subsubsection{The first dipole state \label{Subsec:FDS}} 

\begin{figure}
\includegraphics[angle=-90,width=0.42\textwidth]{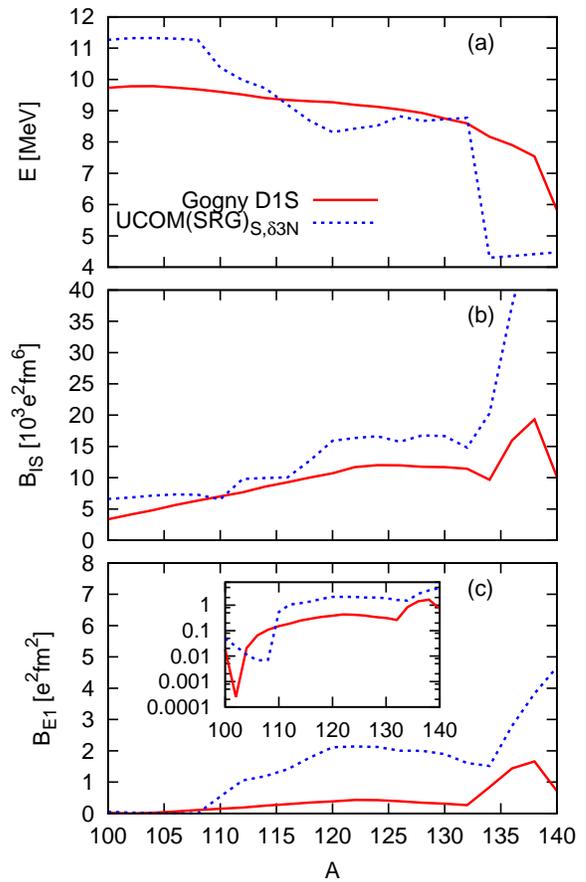}
\caption{
(Color online) Properties of the IS-LED state along the Sn isotopic chain: (a) energy, (b) IS strength and (c) E1 strength. 
The inset shows the E1 srength in logarithmic scale. 
Results are given with the Gogny D1S and the UCOM(SRG)$_{S,\delta 3N}$ interactions. 
\label{Fig:propsISLED}}
\end{figure}

First, we look at the properties of the first and strong IS-LED state. 
In Fig.~\ref{Fig:propsISLED} 
we plot, as a function of mass number $A$ and for both interactions, the energy, IS strength and E1 strength of this state. 
As already observed, these properties vary rather smoothly  
in the case of the Gogny interaction. 
One reason for the less smooth behaviour in the case of the 
UCOM(SRG)$_{S,\delta 3N}$ 
interaction is that there is more than one strong states, while we only plot the properties of the first one. 
Aside from this, we observe that the energy and IS strength is rather similar with both interactions. 
For $^{124}$Sn, IS strength has been observed in $(\alpha,\alpha'\gamma)$ experiments 
shared between two fragmented structures at about 5.5 and 6.5~MeV~\cite{End2010}. 
Our calculations predict a strong IS state at about 8.5-9~MeV and given the above-mentioned energetic shift, 
are in concordance with the experimental findings. 

It is worth noting that the amount of IS strength of the first IS-LED state is similar with both interactions and the values are large, corresponding to a percentage of, 
at the very least, $3\%$, and reaching up to about $10\%$ in the more neutron-rich species, of the total calculated energy-weighted sum of IS strength. 
This finding corroborates the collective character of this state. 
The IS strength has not been extracted from data for any of the Sn isotopes, therefore we cannot compare the calculated IS strength with data. 
However, IS strength of similar magnitude and at the same energy region has been observed in other nuclei~\cite{Poe1992}. 

\begin{figure}
\includegraphics[width=0.47\textwidth]{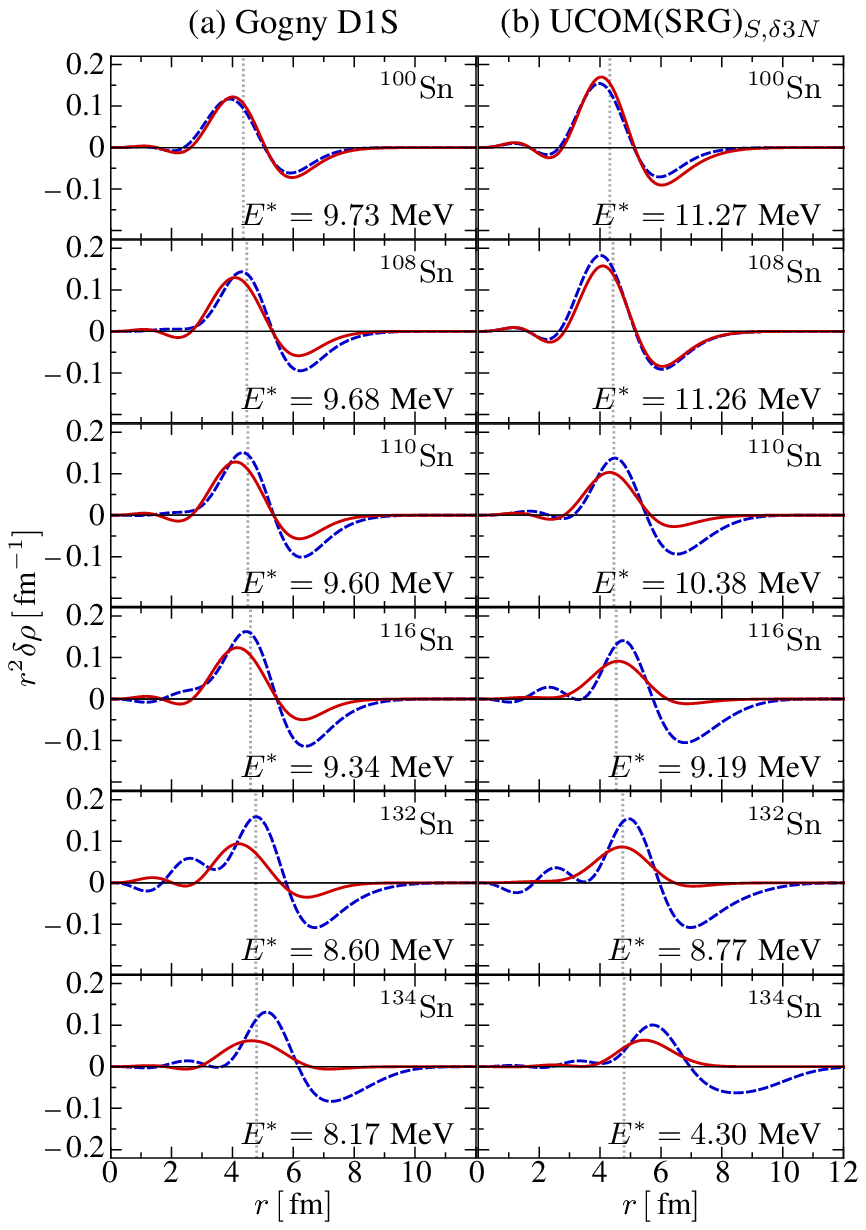}  
\caption{(Color online) For the shown isotopes and interactions, calculated proton (red solid lines) and neutron (dashed blue lines) transition densities 
and excitation energies (as indicated) 
of the IS-LED resonance. 
\label{Fig:TrDen}}
\end{figure}

Before we turn to the E1 strength, it is instructive to inspect the representative transition densities shown in Fig.~\ref{Fig:TrDen}. 
In the $N=Z$ isotope $^{100}$Sn the IS-LED state corresponds to the isospin-forbidden E1 transition studied in detail in Ref.~\cite{PPR2011}. 
In other isotopes the transition densities retain a similar character, namely that of the oscillation of an outer layer of nucleons versus an inner core. 
In general, both protons and neutrons participate in the outer layer. 
Beyond a certain neutron number the neutrons dominate the outer layer and one may speak, at least in a loose sense, of an oscillation of a neutron skin against a core. 
It depends on the model at what neutron number and how smoothly the transition to such a neutron-skin mode takes place. 
For the Gogny D1S interaction, this transition happens very smoothly between the shell closures of $N=50$ and $N=82$, with protons always participating in the outer layer. 
Beyond $N=82$ 
an abrupt transition takes place to a rather pure neutron-skin mode, as illustrated in Fig.~\ref{Fig:TrDen}(a) for $^{134}$Sn. 
At this point the loosely bound neutron hole states of the $pfh$ shell become active. 
The neutron transition densities become more extended.  
In the case of the 
UCOM(SRG)$_{S,\delta 3N}$ 
interaction, 
an abrupt transition from a very IS mode (same transition densities for protons and neutrons) 
to a neutron-dominated mode takes place beyond $N=58$, as shown in Fig.~\ref{Fig:TrDen}. 
The transition densities change quite smoothly beyond that neutron number towards being more neutron dominated. 
Beyond $N=82$ the highest neutron hole states are no longer bound. 

The above differences are reflected on the E1 strengths of the IS-LED, also plotted in Fig.~\ref{Fig:propsISLED}. 
It is interesting to note that the minimum is not reached for the $N=Z$ isotope, but a heavier one. 
In $^{100}$Sn protons slightly dominate in the outer layer, this asymmetry being from the Coulomb interaction~\cite{PPR2011}. 
We also observe an increase of LED strength beyond $^{132}$Sn, in agreement with studies based on Skyrme functionals~\cite{TeE2006,INY201Xa,ENI2013a,AGK2011}. 

The total E1 strength below 9~MeV has been measured in $^{112}$Sn~\cite{Oze2007}, $^{116}$Sn~\cite{Gov1998} and $^{124}$Sn~\cite{Gov1998} 
and it was found to add up to 0.174, 0.204 and 0.345~$e^2$fm$^2$, respectively. 
We notice that the first IS-LED alone, as calculated with the 
UCOM(SRG)$_{S,\delta 3N}$ interaction, 
carries almost 10 times this strength. 
This result, namely that within QRPA the 
UCOM(SRG)$_{S,\delta 3N}$ interaction overestimates the low-energy E1 strength, 
is consistent with the findings of Refs.~\cite{PPR2011,PHP2012}. 
It provides useful guidance to further improvements of the currently very simple prescription used to determine the phenomenological three-body force. 

As in the case of the stable Ca isotopes~\cite{PHP2012}, 
our present results suggest that the first IS-LED state is a universal phenomenon, not particular to $N=Z$ nuclei. 
In the case of $N\approx Z$ nuclei, 
such a state has been produced also by some QRPA and semi-classical calculations based on Skyrme interactions~\cite{TeE2006,Urb2012}, 
but is missing in other models, both relativistic and non-relativistic. 
Generating dipole collectivity at low energies may have to do with the implementation of the (Q)RPA including, as has been speculated before~\cite{TsL2008,PPR2011}, the consistent treatment of the residual spin-orbit force. This is an interesting puzzle that warrants further attention in the future. 

In what follows we will restrict our discussion to the Gogny D1S interaction, which consistently appears in good agreement with the data. 

\subsubsection{Summed E1 strength at low energies} 

We now take a closer look at how our results with the QRPA and the Gogny D1S force compare with existing data at low energies, to validate the model further. 
Summed E1 strength up to almost 9 MeV has been measured in nuclear resonance fluoresence (NRF) experiments for $^{112}$Sn~\cite{Oze2007}, $^{116}$Sn and $^{124}$Sn~\cite{Gov1998}. 
For the unstable isotopes $^{130}$Sn and $^{132}$Sn the amount of strength was measured, with large errorbars, between the particle threshold and the GDR~\cite{Adr2005,Kli2007}. 
 
As already emphasized, one must take an energetic shift into account when comparing QRPA results with data at low energies, 
approximately equal to 3~MeV. 
Once must also keep in mind that fragmentation from effects beyond QRPA may shift a small amount of strength beyond the energy region in which we make comparisons. 
We also note that the strength of individual QRPA states is sensitive to the ingredients of each calculation, precisely because it is very small. 
Finally, the NRF measurements were less precise towards the particle threshold and some E1 strength possibly went undetected. 
For all these reasons, deviations are to be expected. 
 
In Fig~\ref{Fig:sumE1exp} we display the summed E1 strength as calculated within QRPA in various energy regions: 
\begin{figure}
\includegraphics[angle=-90,width=0.42\textwidth]{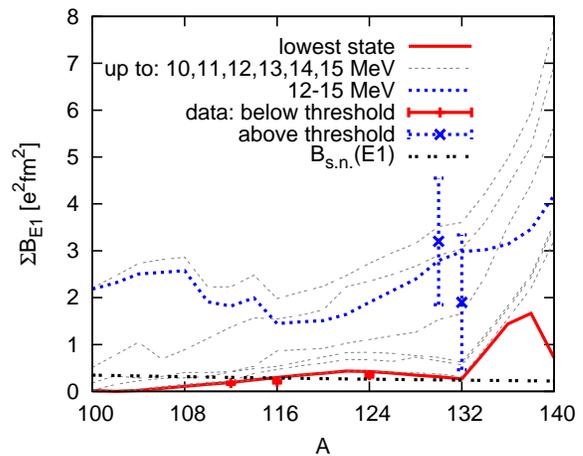}
\caption{(Color online) 
Summed E1 strength within the indicated ranges of excitation energy or of the first state only, 
calculated using the Gogny D1S interaction,  
and measured values below~\cite{Gov1998} or above~\cite{Kli2007} particle threshold. 
The value of one {\em single-neutron} unit, Eq.~(\ref{eq:Bsn}), is also indicated. 
\label{Fig:sumE1exp}}
\end{figure}
\begin{figure*}
\includegraphics[angle=-90,width=0.9\textwidth]{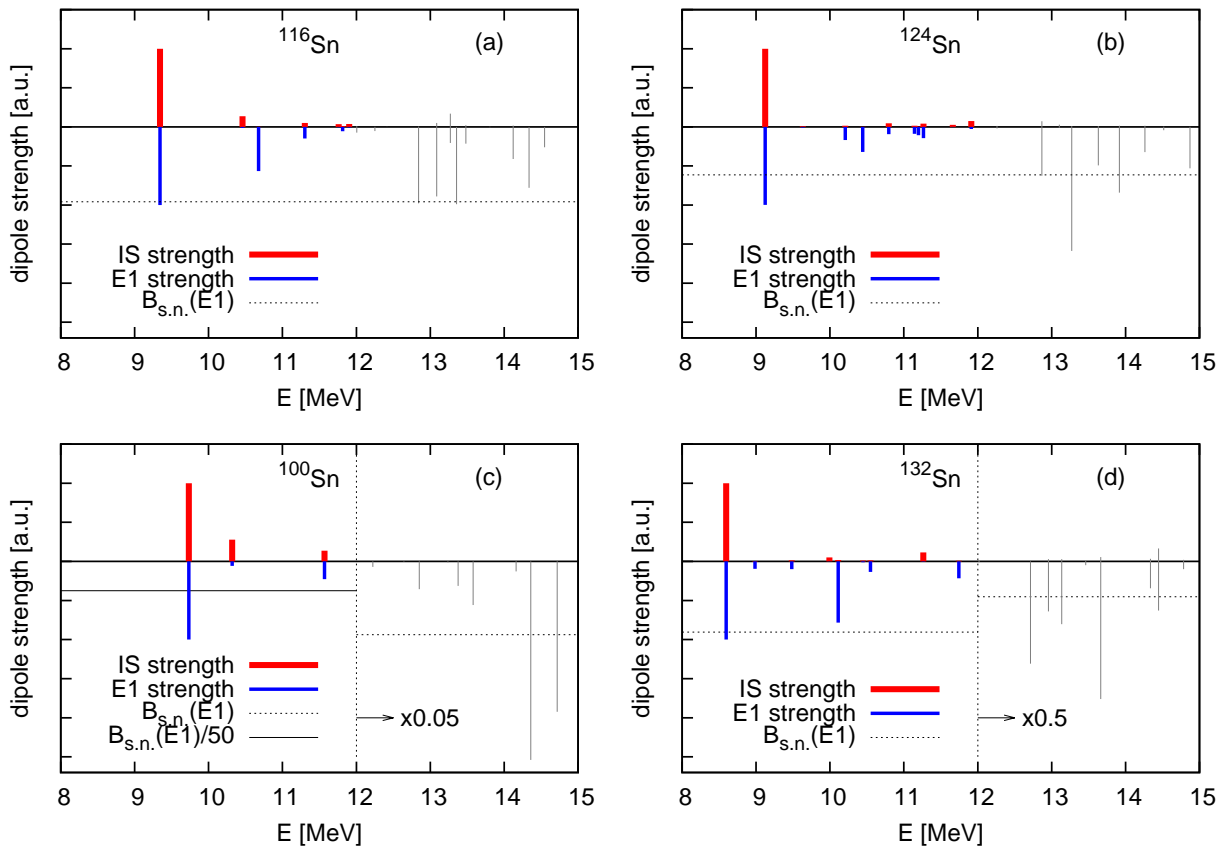}
\caption{(Color online) Low-energy transition strength for the representative isotopes with $A=116,124,100,132$. 
The lowest-energy transition has IS strength of, respectively: 
$9290,12000,3350,11430\, e^2$fm$^6$  
and E1 strength of, respectively: 
$0.297,0.424,0.019,0.264\, e^2$fm$^2$. 
For quantitative comparison, the value of a {\em single-neutron} unit, Eq.~(\ref{eq:Bsn}), is indicated.  
Notice that in the cases of $^{100}$Sn and $^{132}$Sn the scale below and above 12MeV, as marked by the vertical dotted line, is different. 
\label{Fig:IsoSplit}}
\end{figure*}
for the lowest, IS-LED state only; for all states up to 10,~11,~12,~13,~14, or 15~MeV; and for states between 12 and 15~MeV. 
The value of a single-neutron unit as defined by Eq.(\ref{eq:Bsn}) 
is also shown~\footnote{The present values are 3 times smaller than those used in Ref.~\cite{Adr2005}}.  
We observe that the strength measured below threshold ($^{116,124}$Sn) is compatible with the strength of the IS-LED or even, in light of the above discussion, the strength 
calculated up to about 11-12 MeV. 
We find that also the strength measured above threshold ($^{130,132}$Sn, detected at approximately $9-12$~MeV) is compatible with our calculations ($12-15$~MeV). 
This result is very gratifying and lends great confidence to the interpretations and predictions that will follow next. 

An important conclusion that follows from our results is that the strong enhancement of E1 strength measured in the exotic isotopes 
is not due to their large neutron excess, as postulated in Refs.~\cite{Adr2005,Kli2007}, 
but simply due to the energy region accessed, compared to the stable species.  
In fact, we predict that the E1 strength below threshold should be comparable in all Sn isotopes with 
$N\approx 64 - 82$, with much of it coming from states other than the IS-LED mode. 
Moreover, the summed IS dipole strength below threshold should amount to a few per cent of the IS EWSR for {\em all} Sn isotopes with $N=50-82$ 
(generally increasing with $N$ up to $N\approx 126$). 
Last but not least, according to our results a significant amount of E1 strength is expected around threshold and still below the GDR region for all isotopes. 
Regardless of the mechanism that gives rise to this strength, to which we will return in Secs.~\ref{Subsec:Split} and \ref{Subsec:Coll}, its existence alone could be highly consequential for nucleosynthesis processes. 
 
Our predictions for the exhaustion of the classical TRK sum rule and the total energy-weighted IS sum rule (IS-EWSR) in the even $^{100-132}$Sn isotopes, based on the Gogny D1S interaction, are as follows: 
Below threshold no more than, approximately, 1\% of the TRK sum (much less than this value for $^{100}$Sn) and approximately $3-8\%$ of the IS-EWSR is exhausted; 
below 12~MeV (15~MeV calculated energy), in all cases with even $A=100-132$, a few percent (typically $5-10\%$) of the TRK {\em and} the IS-EWSR is exhausted. 
The TRK percentage below 12~MeV would not necessarily vary smoothly with $A$. 
 
The centroid energy of E1 strength below threshold has been reported~\cite{Tof2011} for $^{116,118,122}$Sn as equal to 8.0(1),~8.2(1),~8.6(2), respectively, i.e., rising with $A$. 
Our calculated centroids up to 12~MeV are, respectively, 10.03,~10.02,~9.98~MeV. 
The energy seems to drop very slowly with $A$, in agreement with other theoretical calculations and in disagreement with the data as discussed in ~\cite{Tof2011}. 
Given 
(a) the small amounts of strength involved, (b) the fact that these amounts are spread over several states, (c) that higher-order configurations will result in a redistribution of these small amounts over many more states influencing in unknown ways the centroid (even if the total strength does not change much) and of course d) the unavoidable arbitrariness of the energy cutoff which we have introduced, we are unable to reach a solid conclusion, but we do find likely that a correct description of the centroid energy systematics below threshold is beyond the capacities of (Q)RPA. 
 
\subsubsection{Splitting into different modes\label{Subsec:Split}} 

The isospin structure of LED strength that comes out of our calculations is illustrated in Fig.~\ref{Fig:IsoSplit}, 
for four representative isotopes. 
We see once again that most of the IS strength is carried by the first state, 
but E1 strength appears throughout the low-energy region up to a couple of MeV above the experimentally accessed region in $^{116,124}$Sn (states shown with thin gray lines, above roughly $E=12$~MeV in QRPA). 
Our result is manifestly in line with the observation of isospin splitting of the LED strength in $^{124}$Sn and other heavy nuclei~\cite{Sav2006,End2010,SAZ2013,Der2013}. 

In Fig.~\ref{Fig:IsoSplit} we also indicate the value of the single-neutron unit in each nucleus, Eq.~(\ref{eq:Bsn}), 
keeping in mind that a single-proton unit is $(N/Z)^2$ times that of a neutron unit. 
We observe that most transitions are consistent with single-particle states, leaving no strong  
grounds for expecting collectivity in this channel. 
We will return to this subtle issue in Sec.~\ref{Subsec:Coll}.  

\begin{figure}
\includegraphics[width=0.45\textwidth]{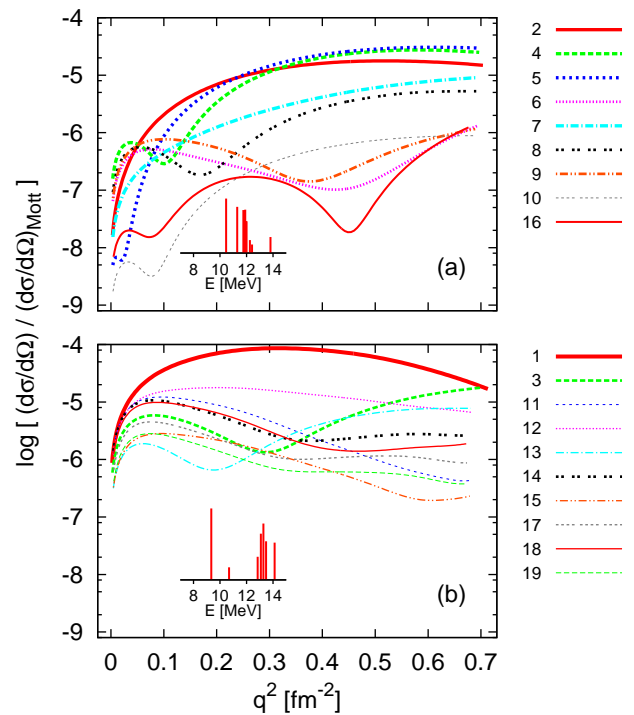}
\caption{(Color online) Longitudinal, electroexcitation form-factor of low-energy dipole states in $^{116}$Sn, 
calculated with the Gogny D1S interaction and within distorted-wabe Born approximation. 
(a) States with $B(E1)<0.5~10^{-2}e^2$fm$^2$. 
(b) States with $B(E1)>0.5~10^{-2}e^2$fm$^2$. 
States are enumerated according to increasing excitation energy. 
Insets: IS strength of respective states, in logarithmic scale and arbitrary units. 
\label{Fig:DWBA}}
\end{figure}

The splitting of the spectrum into different bunches of states is further demonstrated 
in Fig.~\ref{Fig:DWBA}, where we display the longitudinal form factor of the 19 first states of $^{116}$Sn (of energy up to 15~MeV), calculated within the distorted-wave Born approximation (DWBA). 
In Fig.~\ref{Fig:DWBA}a, states with strength $B(E1)<0.5~10^{-2}e^2$fm$^2$ are included. 
In Fig.~\ref{Fig:DWBA}b, all other, stronger, states are included. 
In the respective insets, the IS strength of the respective states is shown, in logarithmic scale and arbitrary units. 
The states are enumerated according to increasing excitation energy. 
We note that states no. 10 and 16 are too weak to be visible in Fig.~\ref{Fig:IsoSplit}.  

Some of the states in Fig.~\ref{Fig:DWBA}a demonstrate a diffraction minimum, reminiscent of that in the IS-LED of $N=Z$ nuclei as well as $^{48}$Ca. 
Interestingly, the first IS-LED state of $^{116}$Sn does not. 
In order to further investigate the kinship of the strong IS-LED modes in various nuclei, a dedicated study of the transverse form factors, accessible in transverse electron scattering, appears worthwhile in future. 


From Fig.~\ref{Fig:DWBA}b it becomes clear that QRPA generates a bunch of eigenstates with similar form factors, non-negligible IS strength and in close proximity energetically (around 13~MeV). In fact, we find that between 12 MeV and the GDR approximately $10\%$ of the IS EWSR is exhausted. This value is clearly lower than the strength detected above threshold in Ref.~\cite{You2004b}, but roughly in agreement with another non-relativistic study~\cite{CVB2000}.  

\subsubsection{Collectivity\label{Subsec:Coll}} 

\begin{figure*}
\includegraphics[angle=-90,width=0.95\textwidth]{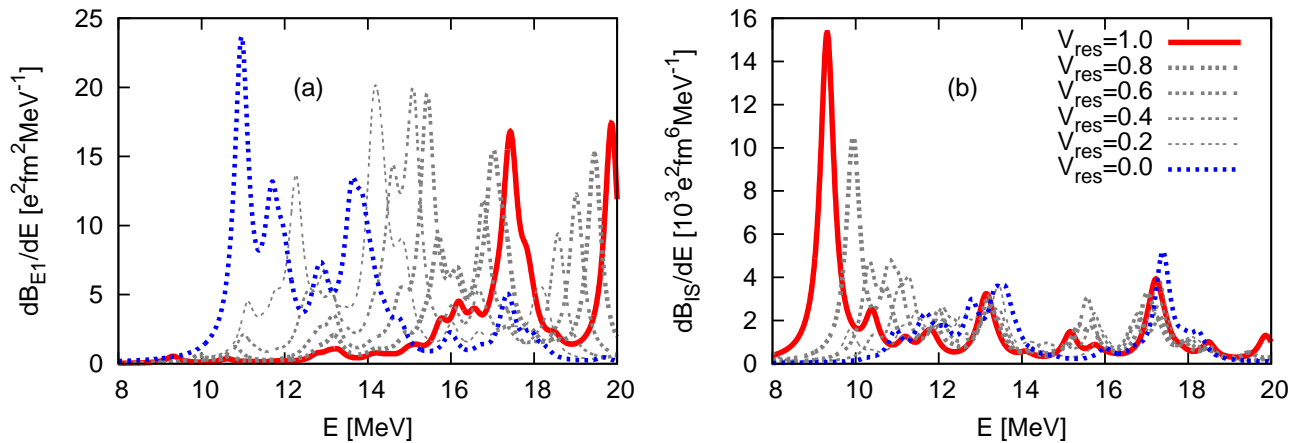}
\caption{(Color online) 
Behavior of the E1 (left) and IS (right) dipole response of $^{116}$Sn under variation of the strength of the residual interaction. 
The Gogny D1S interaction is used. 
The thick, blue dashed lines indicate the transition strength of unperturbed HFB $qp$ transitions, $V_{\mathrm{res}}=0$. 
The red solid lines indicate the full QRPA response, $V_{\mathrm{res}}=1.0$. 
Results corresponding to intermediate values of $V_{\mathrm{res}}$ are also plotted. 
The response is smoothed with a Lorenzian of width equal to 0.2~MeV. 
\label{Fig:vres}}
\end{figure*}
Our attempt to assess the collectivity of E1 strength closer to threshold (in the region of 11-15~MeV in our calculations) with the help of single-neutron particle states was inconclusive. 
The bunching of states discussed above may be indicative of a collective mode, fragmented through the Landau mechanism. 

A conclusive analysis of the  
coherence of dipole states is difficult, because the treatment of the center-of-mass motion introduces subtle ambiguities. 
Indeed, let us begin with the assumption that a transition is 100\% of single-particle (or $2qp$) character. 
This transition entails a finite transition matrix element of the displacement operator. 
This must be compensated by the recoil of the rest of the nucleus, or at the very least another single-particle excitation of opposite displacement. 
Therefore our assumption was wrong and more than one configurations must contribute to any (Q)RPA wavefunction always. 
Of course, if an eigenstate received a large percentage of its norm from a single configuration, we could not argue against its single-particle character. 
Interestingly, we found that hardly any eigenstate is made up by more than $50\%$ by one configuration. 
The majority of eigenstates receive contributions from most $1\hbar\omega$ $2qp$ configurations. 

Regarding the coherence of an excited state in the E1 channel, it makes a difference whether one considers the bare nucleon charges, in which case neutron states do not contribute to the E1 strength {\em at all}, or whether one considers the usual effective charges. 
We stress that the total E1 strength of any given (Q)RPA state (except the spurious one) is not affected by the choice of effective charges, 
but the individual contribution of each single-particle configuration strongly is. 

An alternative way of diagnosing collectivity in QRPA is to compare the QRPA strength function with the unperturbed one. 
We performed this exercise in the case of $^{116}$Sn and show the result in Fig.~\ref{Fig:vres}, for both the IS and E1 strength function. 
We started with the unperturbed spectrum and obtained results for various fractions 
$V_{\mathrm{res}}$ 
of the residual interaction, ranging from 0.2 to 1.0, in which case the full QRPA result is obtained. 

In the IS channel, the collectivity of the first eigenstate is clearly demonstrated. 
This state collects more and more IS strength from higher-lying states, as  
$V_{\mathrm{res}}$ 
increases, while its energy ends up well below the unperturbed spectrum. 
Regarding the E1 strength function, we can clearly see that, as 
$V_{\mathrm{res}}$ 
increases, 
this energy region is depleted of its strength, which appears to be attracted more and more by the higher-lying GDR. 
A relatively small amount remains when $V_{\mathrm{res}}=1$. 
This result implies that, rather than being part of the GDR, the E1 strength in this energy region consists of the remainders of dipole strength that failed to join the GDR, as first suggested in Ref.~\cite{OHD1998} and in other non-relativistic studies of various nuclei~\cite{ReN2013,CDA2013}. 
Nonetheless, together these states carry a non-negligible amount of strength and, as already discussed, could be excited strongly by other operators involving momentum transfer. 
We have checked that similar results are obtained for $^{132}$Sn. 

To summarize, the relatively low IS and E1 strength carried by the individual states close to and above particle threshold speak against their collective character, but the energetic proximity of states with similar form factors supports a collective scenario. 
Clearly, data other than $B(E1)$ strength should become available before we can reach a consensus as to what determines the dipole response in this scarcely explored energy region.

\section{Summary and conclusions\label{sec:con}} 

%
In this work we have studied the low-energy dipole strength distribution along the Sn isotopic chain in both the IS and IV electric channels, 
with the help of the 
self-consistent QRPA with finite-range interactions. 
We have used two types of interaction: the Gogny D1S force and a quasi-realistic interaction consisting of the unitarily transformed Argonne V18 interaction at the two-body level plus a phenomenological three-body contact term. 
We compared our results with existing data. 
As in previous applications, we have found that the quasi-realistic interaction 
overestimates the LED strength. 
Our results with the Gogny interaction are very well compatible with existing data, except that the centroid energy of E1 strength below threshold is predicted to drop with neutron number. This discrepancy is attributed to the limitations of QRPA in describing precisely the fragmentation of strength. 

On the basis of our calculations with the Gogny D1S interaction we have tackled open issues regarding the properties of LED strength, including its collectivity and its so-called isospin splitting, and we have made testable predictions. 
In particular,  
we found that from $N=50$ and up to the $N=82$ shell closure ($^{132}$Sn) the lowest-energy part of the IS-LED spectrum is dominated by a collective transition whose properties vary smoothly with neutron number and which cannot be interpreted as a neutron-skin oscillation. 
For the neutron-rich species this state contributes to the E1 strength below particle threshold, 
but much more E1  
strength is carried by other, weak but numerous transitions around or above threshold. 
We found that strong structural changes in the spectrum take effect beyond $N=82$, namely increased LED strength and lower excitation energies. 
We therefore predict that (a) the summed IS strength  below particle threshold shall be of the same order of magnitude for $N=50-82$, 
(b) the summed E1 strength up to approximately 12~MeV shall be similar for $N=50-82$~MeV, while 
(c) 
the summed E1 strength below threshold shall be of the same order of magnitude for $N\approx 64 - 82$ and much weaker for the lighter, more-symmetric isotopes. 

Our results are in general agreement with other non-relativistic studies~\cite{TeE2006,Urb2012}. 
However, in some studies no collective IS mode is predicted~\footnote{Private communications with P.-G.Reinhard, I.Hamamoto, G.Co'}. 
The discrepancy seemingly has to do with the RPA implementation and not with the interaction used. 
Finally, we observe a possibly radical disagreement with relativistic RPA models, in that  
the latter overestimate the strength below threshold in stable isotopes and predict a collective, neutron-skin excitation at threshold in the unstable isotopes, but no resonance at lower energies~\cite{PVK2007}. 

Experiments in preparation or under analysis will be able to substantially constrain theoretical interpretations of LED strength as well as corroborate or refute the universality of the low-energy IS collective state~\cite{PPR2011,PHP2012} predicted in this work along the Sn isotopic chain.

\pp{(Extended version of abstract. Make sure that the three ``open issues" mentioned in the introduction have been addressed).} 

\begin{acknowledgments}
We are grateful to G.~Co', I.~Hamamoto, and P.-G.~Reinhard for generously sharing their insights and results regarding low-energy coherent states and to T.Aumann for useful remarks. 
This work was supported 
by 
the ANR project ``SN2NS: from supernova to neutron stars and black holes", 
the National Science Foundation under Grant No. PHY-10002478, 
the NUCLEI SciDAC Collaboration under the U.S. Department of Energy Grant No. DE-SC0008533, 
the Deutsche Forschungsgemeinschaft through contract SFB 634, 
the Helmholtz International Center for FAIR (HIC for FAIR), 
the BMBF through contract 06DA7074I, 
and 
by the Rare Isotope Science Project 
of the Institute for Basic Science 
funded by the Ministry of Science, ICT and Future Planning 
and the National Research Foundation of Korea (2013M7A1A1075766). 
\end{acknowledgments}


%

\end{document}